\documentclass[11pt]{article}
\usepackage[margin=1in]{geometry}  % set the margins to 1in on all sides
\usepackage{amsmath}
\usepackage{amssymb}		

\usepackage{pgf}
\usepackage{tikz}
\usetikzlibrary{arrows,shapes,positioning,trees,automata,petri,fit}

\usepackage{graphics}
\usepackage{pgfplots}
\usepackage{multirow}
\usepackage{fancybox}
\usepackage{booktabs}
\usepackage{caption}

\usepackage{hyperref}
\usepackage{url}
\usepackage{wrapfig}
\usepackage{color}

\usepackage[makeroom]{cancel}
\usepackage{setspace}

\newtheorem{prop}{Proposition}

\newtheorem{cor}{Corollary}
\newtheorem{defi}{Definition}

\newtheorem{rem}{Remark}
\newtheorem{ass}{Assumption}
\newtheorem{ex}{Example}

\definecolor{dark_red}{RGB}{200, 0, 0}

\begin{document}

\title{Stable Task Allocation in Multi-Agent Systems with \\ Lexicographic Preferences\footnote{A slightly abridged version of this paper is under review for a journal publication.}}

\author{Spyros Reveliotis and Eva Robillard% <-this % stops a space
\thanks{S.\ Reveliotis is with the School of Industrial \& Systems Engineering, Georgia Institute of Technology, Atlanta, GA, USA, email: {\tt spyros@isye.gatech.edu}.}
\thanks{E.\ Robillard is studying at ENS Paris-Saclay, Universit\'e Paris-Saclay, Gif-sur-Yvette France, email: {\tt eva.robillard@ens-paris-saclay.fr}. This work was carried out while she was a visitor student at Georgia Tech.}
}

\maketitle

\begin{abstract}
Motivated by the increasing interest in the explicit representation and
handling of various ``preference'' structures arising in modern
digital economy, this work introduces a new class of ``one-to-many
stable-matching'' problems where a set of atomic tasks must be stably
allocated to a set of agents. An important characteristic of these stable-matching 
problems is the very arbitrary specification of the task
subsets constituting ``feasible'' allocations for each agent.
It is shown that as long as the agents rank their
feasible task allocations lexicographically with respect to
their stated preferences for each atomic task,
matching stability reduces to the absence of blocking agent-task pairs.
This result, together with a pertinent graphical representation of
feasible allocations, enable (i) the representation of the
space of stable matchings as a set of linear constraints with binary 
variables, and (ii) the specification and handling of certain 
notions of optimality within this space of stable matchings. The last part of the paper also 
addresses the notion of ``substitutability" in the considered problem context.
\end{abstract}

{\bf Keywords:}
Task Allocation in Multi-Agent Systems, Stable-Matching Theory, Cooperative Games

\section{Introduction}
\label{sec:intro}

Some important effects of the current proliferation of the internet-based social and business platforms are (i) the availability of very high volumes of contextual information regarding the discourse  and the various transactions that take place through these platforms to the parties involved, and (ii) the possibility of more directed contacts, ``influencing'' and ``networking'' among these parties based on the available information. Hence, in business-oriented applications of these electronic media, like in e-commerce and entertainment platforms, a prominent theme underlying the corresponding operations has been the effective assessment and exploitation of perceived ``customer preferences'' through various mechanisms and techniques that aim at the interpretation and the influence of the observed behaviors \cite{Foner96,Terveen05}. More recently, there is also an increasing interest in the acquisition, representation and integration of various notions of ``preference'' in the transactions taking place in other internet-facilitated services,  like those taking place in ``ride-sharing'', ``client-server'' computing and certain robotic applications \cite{Dean+06,Wang+18,Prorok+21,Messing+22,MartMori24}. The primary motivation for these recent developments is the realization that the inability to attend to observed or, in some cases, explicitly expressed preferences by the served parties, in view of all the aforementioned visibility of the operational environment that is  enjoyed by them, may lead to their alienation and their eventual disengagement from the corresponding platforms.

Motivated by the above trends and remarks, this work addresses problems of task allocation to a set of available agents under the assumption that, both, the owners of the tasks and the participating agents have explicitly expressed preferences for their potential allocations. A more specific high-level description of the addressed problems that serves as a prototypical motivation of this work, is as follows:

A clearing house allocates a set of geographically distributed tasks to a set of itinerant agents.\footnote{A more concrete example of this task-allocation function is provided, for instance, by the repair services offered in many areas of the United States by a company like Home Depot.}
The tasks enter the system in a dynamical manner and they are allocated to the available agents through a decision process that is activated at regular time intervals (e.g., at the beginning of a working day or every certain number of hours). 
During such an allocation cycle, a task can be assigned to a single agent but an agent can be allocated many tasks. 
More specifically, tasks are classified in categories and each agent has a certain level of qualification (or competence) for performing tasks in each category. In particular, an agent may be unqualified to perform tasks in a certain category, and the corresponding allocations are ``inadmissible''. For admissible task allocations, the agent qualification for the corresponding tasks determines a certain nominal processing time for their execution by this  agent. The allocation of a certain subset of tasks to a certain agent is ``feasible'' if (i) each task in this subset is an admissible task for the considered agent, and (ii) the total time required for the execution of all the allocated tasks and for traveling to the corresponding task locations is within a time budget that is specified by the current allocation cycle.

Furthermore, agents prefer to be allocated tasks for which they are better qualified, and this preference induces a ``preference'' order on the subsets of tasks that constitute feasible allocations for each agent.\footnote{This induced preference relationship for the agent task-allocations is captured by the notion of ``lexicographic preferences'' that qualifies the problems considered in this work. This notion is technically defined later in the paper.}
Similarly, task owners have explicit preferences for the various agents that can be potentially allocated to their tasks; these preferences may be based on prior experience or some {\em a priori\/} communication with these agents, quoted prices, quoted processing times, or, more generally, a combination of many of these factors.

Finally, assuming that the clearing house charges a fixed price for each materialized task allocation, the house objective is to maximize the number of the allocated tasks. But the final overall allocation for each allocation cycle must also be compatible with the expressed preferences by the agents and the task owners. More specifically, in the proposed allocation, there should not exist any set consisting of agents and tasks that would be willing to subvert this allocation by reallocating themselves within the elements of this set in a way that aligns better with their stated preferences; such a subversive set of agents and tasks is characterized as a ``blocking coalition'' for the contemplated allocation, while the allocation itself is considered ``unstable'' and, therefore, undesirable.

It should be clear from the above description that the introduction of the outlined ``preference'' structure in the considered task-allocation problem, and the induced ``stability'' requirement for its solution, differentiate substantially this problem from the task-allocation problems that have been typically studied in the past literature of multi-agent systems, where the desirability -- or the ``value'' -- of the considered allocations are characterized numerically -- and, therefore, more implicitly and frequently more obscurely --  by some ``utility'' function \cite{Haer}. In more specific, technical terms, the posed ``stability'' requirement for the solutions of the task-allocation problems introduced in this work implies that the corresponding formulations must observe a new class of structures and constraints expressing a more qualitative -- or ``ordinal'' -- kind of information regarding the agent and the task preferences, as opposed to the ``cardinal'' type of metrics and constraints that have been considered in the past formulations of such allocation problems.

In formal systems theory, allocation -- or ``matching'' -- problems that observe the satisfaction of explicitly stated preferences by the involved parties, are studied by ``cooperative games'' theory \cite{PelegSud}, and they constitute a particular area of this theory that has come to be known, in more colloquial terms, as ``stable matching'' \cite{GusIrv,RothSot90,Manlovebook}. Hence, the primary objective of this work is to position thoroughly and systematically the task-allocation problems that were outlined in the previous paragraphs in the space of ``stable matching'' problems. With this fundamental understanding, a more detailed listing of the paper objectives and contributions is as follows:

\begin{enumerate}
\item
The paper defines rigorously the considered class of stable-matching problems and positions them carefully in the corresponding literature. This analysis also reveals that the considered stable-matching problems lie at the frontier of the corresponding research, and they are characterized by high levels of representational, analytical and computational complexity.

\item
In view of the aforementioned complexity, it is imperative to investigate systematically the general feasibility and the structural properties of the considered problems, and the preservation of any particular -- or ``special'' -- structure that has enabled the past analytical and computational results of ``stable-matching'' theory. A first important result of this investigation is the existence of problem instances that do not possess any stable matchings, 
and this finding also raises the need to explicitly characterize and control the tolerated instability for these problematic cases.
On the more positive side, it is also shown that  the ``lexicographic'' preferences of the agents for the task subsets that constitute their feasible allocations, enable an efficient characterization and assessment of the stability of any tentative overall allocation by restricting the search for blocking coalitions in the set of admissible (agent, task) pairs.

\item
An additional important development that derives from the lexicographic nature of the agent preferences for their potential allocations, and facilitates a succinct algebraic formulation of the considered problem and its structural analysis, is the encoding of the agent preferences over their feasible task allocations  in a graphical structure that is called the ``lexicographic tree''.

\item
The availability of the developments reported in items (2) and (3) above subsequently have enabled the analytical characterization of the considered task-allocation problems as integer programs (IPs) \cite{Wolsey}. The variables employed by the derived formulations are of a binary nature, an aspect that would place these formulations in the easiest class of IP problems \cite{Wolsey}. However, an additional complication arises from the fact that the number of the employed variables can be exponentially large with respect to the number of the allocated tasks. This issue can be effectively addressed by the application of a customized ``column generation'' technique \cite{Magn} for the solution of the linear programming (LP) relaxations that must be solved in the context of any ``Branch \& Bound (B\&B)''-based solution of the considered IP formulations \cite{Wolsey}. The paper outlines such a column-generation method and the corresponding B\&B implementation for the considered IPs, but due to the imposed page limits, the detailed presentation and assessment of this approach is deferred to a subsequent paper.

\item
On the other hand, the last part of the paper discusses the property of ``substitutability'' for the agent allocations that, when present, ensures the feasibility of the corresponding problem instances and enables the adaptation of existing efficient (polynomial-complexity) ``stable-matching'' algorithms for their solution \cite{RothSot90}. Substitutability is succinctly characterized in the considered problem setting by means of the lexicographic tree, and its practical implications are highlighted through the provision of a set of pointers to related structural and algorithmic developments of the stable-matching theory.
\end{enumerate}

The rest of the paper is organized as follows: The next section provides a brief literature survey of the stable-matching problem that enables the succinct positioning of the considered task-allocation problem in this literature. The survey also highlights some important concepts and results from the theory of stable matching that are significant for the subsequent developments. Section~\ref{sec:m-prob} abstracts the considered task-allocation problems into a new class of stable-matching problems, and defines formally the structure of these problems and all the additional concepts and terminology that are necessary for the further developments of this work. Section~\ref{sec:stab} adapts the notion of ``matching stability'' to the considered class of matching problems and analyzes its properties. Section~\ref{sec:IP} presents the IP-based representation of the set of stable matchings, and discusses how this representation can be employed for the 
definition and resolution of various notions of ``optimality'' over this set. Section~\ref{sec:sub} considers the notion of ``substitutability'' in the context of the considered problems and its structural and computational implications. Finally, Section~\ref{sec:con} concludes the paper and suggests some directions for future work.

\section{Literature review of the stable-matching problem}
\label{sec:lit-rev}

The \textit{stable-matching} problem was first introduced through the \textit{Stable-Marriage (SM)} problem in the seminal paper of Gale and Shapley in 1962 \cite{GaleShap62}. Informally, a \textit{stable marriage} is a one-to-one pairing of two sets -- the men and the women -- in which no man-woman pair would agree to leave their assigned partners in order to marry each other. A fundamental result of \cite{GaleShap62} is a very efficient algorithm for the computation of a stable matching, known as the {\em Gale-Shapley (GS)\/} algorithm. Furthermore, the algorithm dynamics and its correctness-analysis provide a constructive proof for the {\em feasibility\/} of every instance of the \textit{SM} problem.\footnote{i.e., of the fact that every SM problem instance admits at least one stable matching.} 
Subsequent publications on the {\em SM\/} problem unveiled deep structural relationships for the underlying solution space, and employed the obtained results in various practical applications. The monograph by Gusfield and Irving \cite{GusIrv}, published in 1989, collects the major developments on the \textit{SM} problem by the time of its publication, and it constitutes a standard reference on the structural and the algorithmic aspects of the corresponding theory. 

The work of Donald Knuth \cite{Knuth76} is another seminal development focusing on the mathematical analysis of matching algorithms for the \textit{SM} problem. One of the most important developments in \cite{Knuth76} is the establishment of a distributive-lattice structure among the stable matches of any \textit{SM} problem instance that aligns with the atomic preferences of the men (or, symmetrically, the women). This structure is also reflected in the solutions that are provided by the {\em GS\/} algorithm, and it enables a systematic and efficient parsing of the underlying solution space.

Along with the one-to-one Stable-Marriage problem, Gale and Shapley introduced the \textit{College Admissions} problem, a one-to-many generalization of the \textit{SM} problem \cite{GaleShap62}. This model is equivalent in structure to another one-to-many generalization known as the \textit{Hospital/Residents (HR)} problem, which will be our primary focus for reference. This problem describes the labor market for medical interns, in which, every year, the hospitals seek to hire a set of well-qualified residents in order to fill a number of available positions, and the residents seek to be employed by their preferred hospital. The \textit{HR} problem was first explored by Roth in \cite{roth1984evolution}. Roth's work also revealed significant structural differences between the {\em SM\/} and the {\em HR\/} problems, and provided a game-theoretic analysis for these stable-matching problems that raises interesting strategic questions regarding the truthful statement of the preferences of the various players \cite{roth1985college}. A comprehensive account of all these developments, plus additional extensions and results on bipartite stable-matching problems, is provided in the monograph by Roth and Oliveira Sotomayor \cite{RothSot90}. Among the developments of \cite{RothSot90}, some of particular interest to this work are (i) the provision of {\em GS\/}-type of algorithms for the {\em HR\/} problem,
(ii) the extension of the distributive-lattice structure of the {\em SM\/} problem, and many of its more practical implications, to the {\em HR\/} case, and (iii) the {\em ``Rural Hospitals''\/} theorem; this theorem establishes a consistency across all the stable matchings for any given {\em HR\/} problem instance with respect to (a) the entire set of the matched residents across all hospitals and (b) the assignments received by each individual hospital, in terms of the number and, in certain cases, even the composition of the sets of their assignees.

The aforementioned developments on the {\em HR} problem have found extensive practical applicability in the context of 
the National Resident Matching Program (NRMP) in the USA, as well as in other centralized automated matching schemes that allocate graduating medical students (residents) to hospital posts. Furthermore, the existence of \textit{couples} who wish to be located at the same hospital, or at hospitals that are geographically close, has given rise to an important variant of the \textit{HR} problem known as the \textit{Hospital/Residents problem with Couples (HRC)} \cite{roth1984evolution, ronn1990np, GusIrv, roth1992two}. In the particular case where every couple seeks to be placed at the same hospital, both, single residents and couples can be treated as {\em atomic entities\/} which are, respectively, requesting one or two positions at their target hospitals. This case is treated and generalized in \cite{mcdermid2010keeping}, which investigated the \textit{Hospitals/Residents problem with Sizes (HRS)}. Some additional studies of this last problem are provided in \cite{Dean+06,CsehDean16}. These two works also broaden the application scope of the {\em HRS\/} model to some further classical problems studied by Operations Management (OM) under the banner of \textit{stable allocation} or \textit{ordinal transportation} problems. 

Three important attributes of the {\em HRC\/} and the {\em HRS\/} problems are that (i) an instance of these two problems need not admit a stable matching \cite{roth1984evolution}, (ii) their solution space does not possess a lattice structure similar to the corresponding structure established for the {\em SM\/} and the basic {\em HR\/} problems, and (iii) the ``Rural Hospitals'' theorem does not hold anymore. Also, 
the problem of deciding whether an \textit{HRC} or {\em HRS\/} problem instance admits a stable matching is strongly NP-complete, even for problem variations that restrict the resident sizes and the hospital capacities to some very small values \cite{ronn1990np,  mcdermid2010keeping, biro2014hospitals}. 

On the more positive side, the works of \cite{BaiouBal00,BaiouBal02} have established that the {\em HRS} problem retains all the important structural and computational / algorithmic properties of the basic {\em HR\/} problem if we allow a partial allocation of the residents with sizes greater than one to more than one hospitals. Moreover, similar results can be obtained for a further generalization of this problem to a many-to-many setting and to problem variations involving real-valued sizes. 

For the harder {\em ``unsplittable''\/} -- or {\em  inseparable\/} -- cases of the {\em HRC} and {\em HRS\/} problems, the literature provides methodology, in the form of {\em Integer Programming (IP)\/} and {\em Constraint Programming (CP)\/} formulations, for assessing the feasibility of any given problem instance, and for computing an optimized stable matching according to certain performance criteria \cite{biro2014hospitals,Manlovebook}. In addition, in the case of infeasible problem instances, these formulations are modified to allow for controlled levels of instability \cite{biro2014hospitals,Wang+18} as a means for coping with the problem infeasibility. An alternative mechanism for coping with the potential infeasibility of the {\em HRS\/} problem has been pursued in the works of \cite{Dean+06,CsehDean16}, which (re-)establish a notion of ``feasibility'' by relaxing -- or ``softening'' -- the capacity constraints of the hospitals. By adapting to the ``unsplittable'' {\em HRS\/} problem some results originally developed by Shmoys and Tardos for the {\em unsplittable transportation problem\/} \cite{shmoys1993scheduling}, the authors of \cite{Dean+06,CsehDean16} were able to provide computational algorithms for the ``unsplittable'' {\em HRS\/} problem that are similar in spirit and complexity with the GS algorithm of \cite{GaleShap62} and guarantee a very tight bound for the capacity violations incurred by the returned solution.

The literature also avails of many additional results for the stable-matching problems outlined in this section and some variations of them, of extensive theoretical and practical interest. An excellent exponent of a very large part of this material is the more recent monograph by Manlove \cite{Manlovebook}, which also considers ``matching under preference'' problem versions that  go beyond the bipartite stable-matching problems considered in this work.\footnote{The perusal of \cite{Manlovebook} also reveals the focal differences between those groups in Computer Science (CS) and Operations Research (OR) that work on the computational analysis and the development of efficient algorithms for the considered matching problems, and the groups coming from the communities of Economics (ECON) and other social sciences that analyze the game-theoretic structures and the strategizing properties of the considered problems and their economic interpretations.}

We also want to point out that, among all the considered developments, the work of Michel Balinski and his collaborators reveals and pronounces the significance of a clever use of graphs in the representation and the analysis of stable-matching problems. The graphical representations that have been employed by these authors, have enabled some very original and very insightful paths for the analysis of the stable-matching problems addressed in those works, and the development of some very elegant solution approaches  \cite{BalRat97,BaiouBal00b}. They have also provided inspiration for our own work.
 
Finally, in the scope of all the aforementioned developments on the stable-matching problem, the task-allocation problem addressed in this work can be perceived as the extension of the {\em HR} problem with sizes to the more general case where the {\em feasibility\/} of the allocation of a set of residents to a hospital is defined by a much more arbitrary -- and less algebraically concise -- criterion than the mere fact that the total number of positions requested by the contemplated allocation must respect the number of available positions announced by the hospital. As pointed out in the introductory section, and further corroborated by the above discussion, this realization places the considered task-allocation problem at the frontier of the current stable-matching theory. It also reveals the high representational and computational complexity of the analysis of this new class of stable-matching problems in terms of: (i) characterizing the solution space of these problems -- i.e., defining the notion of stable task-allocations; (ii) assessing the feasibility of their various instances -- i.e., resolving the existence of stable task-allocations for each problem instance; and (iii) obtaining such a stable task-allocation, if it exists. At the same time, the affinity of the task-allocation problems addressed in this work to the stable-matching problems and their current theory outlined in the previous paragraphs, provide a rich and solid technical base for their systematic investigation. The rigorous  and thorough establishment of this base is the primary contribution of this work.

\section{The considered matching problem}
\label{sec:m-prob}

This section abstracts the task-allocation problem outlined in the introductory section into a novel stable-matching problem, defines thoroughly the structure of this problem, and introduces a graphical representation of certain problem elements that will facilitate the subsequent analysis.

The considered stable-matching problem is a {\em bipartite one-to-many\/} matching problem defined over two disjoint finite sets $A$ and $T$. Set $A$ is a set of \textbf{agents} and set $T$ is a set of \textbf{tasks}. Each agent may be matched with multiple tasks, but each task may be matched with at most one agent.
%Also, in the following, we assume that $|A| = m$ and $|T| = n$, with $m, n \in \mathbb{N}$.

{\bf Acceptability and Preference:}
From the sets of tasks and agents, we define a set $\mathcal{E} \subseteq A \times T$ of \textbf{acceptable} agent-task pairs. 
A pair $(a,t)\in \mathcal{E}$ implies that agent $a$ is qualified for executing task $t$. Pairs $(a,t)$ not in $\mathcal{E}$ cannot occur in the considered matchings.

Set $\mathcal{E}$ induces for each agent $a \in A$ an \textbf{acceptable} set of tasks $T(a)$, defined by 
\begin{equation}
    T(a)= \{t \in T : (a,t) \in \mathcal{E}\}
\end{equation}
Similarly, each task $t \in T$ has an \textbf{acceptable} set of agents $A(t)$, such that 
\begin{equation}
    A(t)= \{a \in A : (a,t) \in \mathcal{E}\}
\end{equation}
The above definitions further imply that for every pair $(a,t)\in A\times T$,
\begin{equation}
t\in T(a)\ \Longleftrightarrow\ a\in A(t)
\end{equation}

For each element $t \in T$ (resp., $a\in A$), we define a \textbf{preference list} in which set $A(t)$ (resp., $T(a)$) is ranked in strict order.
 Furthermore, to avoid an overloading of the notation, in the following, $A(t),\ t\in T$, (resp., $T(a),\ a\in A$) will denote both the unordered sets and their ordered versions. The preferences communicated by each preference list $A(t), t\in T$, represent the choice task $t$ will make if it is faced with a choice between two agents; more specifically, given two agents $a,a'\in A(t)$, task $t$ is said to \textbf{prefer} $a$ to $a'$ if $a$ precedes $a'$ in $A(t)$. The \textbf{preference} relation is defined similarly for agent $a$. These preferences are formally denoted as follows:
    \begin{equation}
        t \textbf{ prefers } a \text{ to } a' ~~\Longleftrightarrow ~~a \underset{t}{>} a'
    \end{equation}
    \begin{equation}
        a \textbf{ prefers } t \text{ to } t' ~~\Longleftrightarrow ~~t \underset{a}{>} t'
    \end{equation}

We further assume that a task prefers to be allocated to an acceptable agent than remain unassigned. Similarly, an agent prefers any singleton allocation to it than no allocation at all. With a slight abuse of notation, we formally denote these facts as follows:\footnote{
The main developments of this work extend to the case where the considered preference lists define only a {\em partial\/} order for the elements of the sets $A(t)$ and $T(a)$; i.e., to the case where the tasks (resp., agents) might be indifferent regarding the choice among a subset of agents (resp., tasks). Since the corresponding adjustments are rather straightforward, we are focusing on total orders for the sake of the simplicity of the pursued presentation.}
\begin{equation}
        \forall t\in T,\ \forall a\in A(t),\ \ a \underset{t}{>} \emptyset
        \label{eq:ab-1}
    \end{equation}
    \begin{equation}
      \forall a\in A,\ \forall t\in T(a),\ \   t \underset{a}{>} \emptyset
      \label{eq:ab-2}
    \end{equation}

{\bf Feasible task allocations:}
When assigning multiple tasks to a single agent, it is essential to identify which subsets of tasks can be assigned simultaneously to it. The sets of tasks that can be allocated to an agent $a$ are referred to as its \textbf{feasible allocations\/}. Every agent $a$ has its own feasible allocations. This fact is formalized in the following definition.

\begin{defi}
\label{defi:feas}
    For every agent $a\in A$, there exist $\nu_a$ subsets of tasks, $S_i^a \subseteq T(a), i \in \{1,\ldots, \nu_a\}, $ whose allocation to $a$ is feasible. The set of all feasible subsets of tasks for an agent $a$ is denoted by $F(a)$. Formally,
\begin{equation}
    \forall a \in A,
    ~~~~F(a) =\big \{S_i^a \subseteq T(a) : i \in \{1,\ldots,\nu_a \} \big \}
\end{equation}
\end{defi}

Sets $F(a),\ a\in A$, are the {\bf feasibility sets} for the corresponding agents. In this work, these sets are further qualified as follows:
\begin{ass} 
\label{ass:down-closed}
For each agent $a \in A$, we assume that the feasibility set $F(a)$ is \textbf{downward closed}; i.e., if $S' \subset S\subseteq T(a)$ and $S \in F(a)$, then $S'  \in F(a)$. 
\end{ass}

Assumption~\ref{ass:down-closed} implies that $\emptyset \in F(a),\ \forall a\in A$. It also implies that each set $F(a)$ can be represented more compactly through its subset $\widetilde{F}(a)$ containing only its {\em maximal elements\/}, where maximality is considered with respect to set closure.
Furthermore, in various parts of the subsequent developments, each task set $S^a_i = \{t_1, t_2,\ldots,t_l\}$ is listed in decreasing task preference according to the corresponding task preference list $T(a)$, and it is represented by the sequence $t_1 t_2\ldots t_l$.

\begin{rem}
\label{rem:HRS}
The feasibility condition encoded by the sets $F(a)$ and Assumption~\ref{ass:down-closed} enables the modeling of ``budget"-type of constraints, like the available positions in the {\em HRS\/} problem discussed in Section~\ref{sec:lit-rev} or the processing-time budgets considered in \cite{Dean+06,CsehDean16}. In both of these cases, each agent $a$ possesses a {\bf capacity\/} $C(a)$, and the assignment of a task $t$ to agent $a$ consumes an amount $s(t;a)$ from this capacity; the latter is known as the {\bf size\/} of task $t$ with respect to agent $a$. 
Hence, in these cases, membership of a set $S\subseteq T(a)$ in the corresponding set $F(a)$ can be easily tested through the condition 
%a by ``knapsack''-type of constraint \cite{papSteigl} ensuring that the corresponding allocation does not exceed the agent availability (or {\bf capacity})
\begin{equation}
\sum_{t\in S} s(t;a) \leq C(a)
\label{eq:HRS-feas}
\end{equation}

But the characterization of a feasible allocation in Definition~\ref{defi:feas} also allows for more general situations where membership in $F(a)$ cannot be characterized by a ``knapsack''-type of constraint like the inequality of Eq.~\ref{eq:HRS-feas}.
In the context of the task-allocation problem that was outlined in the introductory section, the inability to characterize  the feasibility of the allocation of a task subset $S\subseteq T(a)$ to an agent $a$ based on some available time budget $C(a)$, arises from the need to account for the agent traveling time to the locations of its assigned tasks. In general, this traveling time will depend upon the sequence that the agent visits the task locations, and, in an optimized setting, it even requires the formulation and solution of a Traveling Salesman Problem (TSP) \cite{Magn}. A similar effect can arise in the task allocations considered in \cite{Dean+06,CsehDean16} in the presence of sequence-dependent setups \cite{Pinedo}. 
In the rest of this work, we assume that the membership in $F(a)$ of any set $S\subseteq T(a)$ is based on some criterion that satisfies Assumption~\ref{ass:down-closed}, and it is
resolved by means of an {\em ``oracle''\/}.\footnote{\label{foot:oracle} Besides establishing a generic perspective on the logic that defines the feasibility of the considered task allocations, this assumption also implies that we are treating the decision problem `$S\in F(a)\ ?$', for any $S \in T(a)$, as an $O(1)$ task.}
\end{rem}

{\bf The lexicographic ordering of task sets:}
Next, we impose an ordering on the elements of the feasibility sets $F(a),\ a\in A$, that expresses the agent preferences for the corresponding task allocations. This ordering is the \textbf{lexicographic ordering} of each set $F(a)$ that is induced by agent $a$'s preference list for its acceptable atomic tasks, $T(a)$.
%\footnote{The proposed lexicographic ordering is consistent with the fact that in the preference lists $T(a)$, the tasks $t$ appear in decreasing preference.}
A complete formal characterization of this ordering of the set $F(a)$ is as follows:

\begin{defi}\label{def:order}
    Given two elements of the set $F(a)$, $S_1=t_1\dots t_i\dots t_m$ and $S_2=t_1'\dots t_i' \dots t_n'$, let $i\in \big \{1,\ldots,\max(m, n)\big \}$ be the first position where these two sequences differ. We define
    \begin{equation}
        S_1 \underset{a}{>} S_2 \ \ \Longleftrightarrow \ \
          t_i \underset{a}{>}  t_i'
    \label{eq:ord}
    \end{equation}
\end{defi}

A complete interpretation of Equation~\ref{eq:ord} must also consider the convention of Equation~\ref{eq:ab-2}. This  equation implies that a string $t_1t_2\ldots t_n$ is preferred to any of its prefixes. 
In the following, we denote by $P(a)$ the ordered version of the set $F(a)$ according to this lexicographic ordering. 
The reader should also notice that $P(a)$ is a {\em total\/} order.

%\begin{rem}
%A practical justification for the lexicographic preferences established by Definition~\ref{def:order} is the observation that, in many applications, matchings are updated dynamically to account for the emergence of new tasks or for various contingencies in the overall operation of the underlying system, and this continual revision of the agent assignments motivates a {\em ``greedy''\/} attitude from their side with respect to the value that they perceive in each single atomic task.
%\end{rem}

{\bf The lexicographic tree:}
For each agent $a \in A$, the list $P(a)$ can be represented graphically as a {\em rooted tree\/}, to be called the \textbf{lexicographic tree} and denoted by $\mathcal{D}(a)$. The structure of the tree $\mathcal{D}(a)$ is as follows: 
\begin{itemize}
    \item The root node of tree $\mathcal{D}(a)$ is denoted by $o$ and constitutes level 0 of the tree.
    \item Each node $q$ of the tree $\mathcal{D}(a)$, other than the root node $o$, is labeled by a task $t\in T(a)$. In the following, we shall say that node $q$ {\bf holds} task $t$. 
    \item The first level of nodes in $\mathcal{D}(a)$ holds the atomic tasks $t\in T(a)$, arranged from left to right in decreasing preference of their held tasks, according to the task preference list $T(a)$.
    \item For each first-level node $q$ of $\mathcal{D}(a)$, holding some task $t^1 \in T(a)$, and for every task $t^2 \in T(a)$ for which $t^1t^2 \in F(a)$, there is a child of $q$, $q'$, holding task $t^2$. Furthermore, the children of $q$ are arranged from left to right in decreasing preference of their held tasks, according to the task preference list $T(a)$. 
    \item For every node $q$ of $\mathcal{D}(a)$ at a level $n\geq 2$, holding some task $t^n \in T(a)$,  let $t^1t^2\dots t^n$ denote the sequence of tasks held by the nodes on the path that leads from root node $o$ to the considered node $q$. Then, for every task $t^{n+1} \in T(a)$ for which $t^1t^2\dots t^nt^{n+1} \in F(a)$, node $q$ has a child holding $t^{n+1}$. Furthermore, the children of $q$ are arranged from left to right 
 in decreasing preference of their held tasks, according to the task preference list $T(a)$.
\end{itemize}

For every node $q$ of $\mathcal{D}(a)$, we denote the path leading from the root node $o$ to node $q$ by $\pi(q)$. Also, $T(q)$ denotes the set of tasks held by the nodes on path $\pi(q)$. Task set $T(q)$ is a feasible allocation for agent $a$, that is represented uniquely by node $q$. We also set $T(o) = \emptyset$. 
Then, for a node $q$ in some level $i$ of tree $\mathcal{D}(a)$, for $i=0,1,2,\ldots$, $|T(q)|=i$. 

Next, consider two paths $\pi$ and $\pi'$ of tree $\mathcal{D}(a)$ leading from the root node $o$ to the nodes $q$ and $q'$ in the same layer $i$ of tree $\mathcal{D}(a)$. Furthermore, suppose that node $q$ is to the left of node $q'$ within layer $i$. A simple inductive argument can establish that task allocation $T(q)$ is preferred by agent $a$ over the allocation $T(q')$. Similarly, it can be shown that the task allocation $T(q)$ corresponding to a node $q$ of $\mathcal{D}(a)$ is preferred by agent $a$ over the allocation $T(q')$ corresponding to an interior node $q'$ of the path $\pi(q)$. Hence, the most preferred allocation in $P(a)$ is that represented by the node $q^*$ of $\mathcal{D}(a)$ which is its {\em leftmost leaf node\/}. 

In the following, we denote the number of nodes in tree $\mathcal{D}(a)$ by $|\mathcal{D}(a)|$, and we treat this quantity as a measure of the {\em ``size''\/} of this data structure.
We also assume that the nodes $q$ of tree $\mathcal{D}(a)$ are numbered in decreasing preference by agent $a$ of the corresponding allocations $T(q)$; hence, node $q^*$ is numbered by `1', and the root node $o$ is numbered by $|\mathcal{D}(a)|$.

\begin{ex}
\label{ex:l-t}
We demonstrate the various concepts introduced in the previous parts of this section through a small example with agent set $A = \{a_1,a_2\}$ and task set $T=\{t_1,t_2\}$. The set of acceptable pairs ${\cal E}$ is equal to $A\times T$, and therefore, $T(a_1) = T(a_2) = T$ and $A(t_1) = A(t_2) = A$. The preferences of tasks $t_1$ and $t_2$ for the two agents are respectively defined by the strings $a_2a_1$ and $a_1a_2$. Similarly, the preferences of agents $a_1$ and $a_2$ for their acceptable tasks are respectively defined by the strings $t_1t_2$ and $t_2t_1$. The 
maximal task allocations for agents $a_1$ and $a_2$ are, respectively, $\widetilde{F}(a_1) = \{t_1t_2\}$ and $\widetilde{F}(a_2) = \{t_1,t_2\}$. 

\begin{figure}[t]
    \centering
      \begin{tikzpicture}[level/.style={sibling distance=20mm/#1}]
\node [circle,draw,label=left:4] (z){$ $}
  child {node [circle,draw,label=left:2] (a) {$t_1$}
    child {node [circle,draw,label=left:1] (b1) {$t_2$}}}
  child {node [circle,draw,label=left:3] (b) {$t_2$}
  child [grow=right] {node (1) {Level 1} edge from parent[draw=none]
          child [grow=up] {node[rectangle,draw]  (0) {\textbf{$\mathcal{D}(a_1)$}} edge from parent[draw=none]
            child [grow=down] {node (nn) {$ $} edge from parent[draw=none]
            child [grow=down] {node (2) {Level 2} edge from parent[draw=none]}}
          }
    }} 
;
\end{tikzpicture}
%\hspace{1cm}
\vskip 0.1in
      \begin{tikzpicture}[level/.style={sibling distance=20mm/#1}]
\node [circle,draw,label=left:3] (z){$ $}
  child {node [circle,draw,label=left:1] (a) {$t_2$}}
  child {node [circle,draw,label=left:2] (c) {$t_1$}
  child [grow=right] {node (1) {Level 1} edge from parent[draw=none]
          child [grow=up] {node[rectangle,draw]  (0) {\textbf{$\mathcal{D}(a_2)$}} edge from parent[draw=none]
            %child [grow=down] {node (nn) {$ $} edge from parent[draw=none]
            %child [grow=down] {node (2) {Level 2} edge from parent[draw=none]}}
          }
    }} 
  ;
\end{tikzpicture}
\caption{The lexicographic trees of Example~\ref{ex:l-t}.}
\label{fig:lex-trees}
\end{figure}
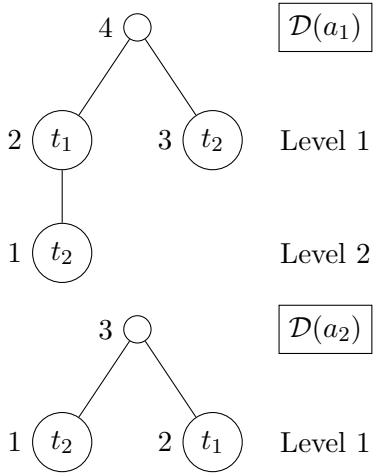

From the above information, we obtain the lexicographic trees ${\cal D}(a_1)$ and ${\cal D}(a_2)$ depicted in Figure~\ref{fig:lex-trees}. As we can see in this figure, when accounting for the possibility of being unmatched, agent $a_1$ has four possible allocations and agent $a_2$ has three. These are the sets $T(q)$ corresponding to each node of the trees ${\cal D}(a_1)$ and ${\cal D}(a_2)$. Also, the numbers next to each node $q$ rank the agent's preference for the corresponding set $T(q)$. Finally, it is worth-noticing that each tree ${\cal D}(a_i),\ i=1,2$, encodes the information regarding the corresponding acceptable-task set $T(a_i)$ and agent's $i$ preferences over this set, in its first layer.
\end{ex}

{\bf Complexity considerations:}
The above characterization of the informational content of the lexicographic tree $\mathcal{D}(a)$ implies that the number of nodes at each level $i$ of this tree is $O\big ( \left (\begin{tabular}{c} $|T(a)|$\\ $i$ \end{tabular} \right ) \big )$. Hence, the tree size $|\mathcal{D}(a)|$ is a super-polynomial function of the number of tasks, $|T|$. 
In many practical cases, the actual size of the corresponding trees $\mathcal{D}(a)$ will be significantly controlled by the imposed feasibility conditions that will restrict the feasible allocations to task sets of small cardinality. But, in general, 
the number of nodes in $\mathcal{D}(a)$ can grow to very large values even for a moderately sized  acceptable-task set $T(a)$. 

This work uses the considered trees $\mathcal{D}(a)$ primarily in order to characterize certain concepts and properties for the considered stable-matching problem. Furthermore, when it comes to developments of a more computational nature, the trees $\mathcal{D}(a)$ may be used in order to guide and localize some iterative  computations that take place in the context of a broader algorithm. These uses of the trees ${\cal D}(a)$, and their significance, are further substantiated and demonstrated in the developments of the following sections.

{\bf Assignments and matchings:}
Following \cite{Manlovebook}, we define an \textbf{assignment} $\mathcal{A}$ as any subset of $\mathcal{E}$, the set of acceptable agent-task pairs. If $(a,t) \in \mathcal{A}$, $a$ is \textbf{assigned} to $t$, and $t$ is \textbf{assigned} to $a$. The set of assignees of an element $x \in A \cup T$ under $\mathcal{A}$ is denoted by $\mathcal{A}(x)$. Also, $x \in A \cup T$ is said to be \textbf{unassigned} if $\mathcal{A}(x)=\emptyset$.

A {\bf matching\/} $\mathcal{M}$ is an assignment satisfying some additional structural and feasibility conditions, and it is formally defined as follows: 
 
\begin{defi}
\label{defi:match}
    A \textbf{matching} $\mathcal{M}$ is an assignment such that 
\begin{enumerate}
    \item $\forall t\in T,\ |\mathcal{M}(t)|\leq 1$.
\item $\forall a\in A,\ \mathcal{M}(a) \in F(a)$.
\end{enumerate}
\end{defi}

We shall refer to $\mathcal{M}(x),\ x\in A\cup T$, as the \textbf{match} of $x$ in $\mathcal{M}$. A matching $\mathcal{M}$ can also be defined by specifying the matches for the assigned tasks and agents in it. In particular, $\mathcal{M}$ can be thoroughly defined by specifying the matches $\mathcal{M}(t)$ for all tasks $t\in T$ or the matches $\mathcal{M}(a)$ for all agents $a\in A$. We also notice that Condition 1 of Definition~\ref{defi:match} further implies that for any two agents $a$, $a'$, $\mathcal{M}(a) \cap \mathcal{M}(a') = \emptyset$.

Furthermore, in the following, the set of all possible assignments for a given problem instance is denoted by $\mathbb{A}$, and the set of all possible matchings is denoted by $\mathbb{M}$.

{\bf Matching domination:}
Next, we introduce some {\bf ordering relations\/} over the elements of $\mathbb{M}$.
%\begin{definition} \label{def: match-order}
    For any two matchings $\mathcal{M}$ and $\mathcal{M}'$, we define:
    \begin{enumerate}
    \item the {\bf task-preference} of $\mathcal{M}$ over $\mathcal{M}'$ by
    \begin{eqnarray}
        \mathcal{M} \underset{T}{\geq} \mathcal{M}' & \Longleftrightarrow \nonumber \\
         \forall t \in T,\ \mathcal{M}(t) \underset{t}{>} \mathcal{M}'(t)~ \vee ~\mathcal{M}(t) = \mathcal{M}'(t)
    \end{eqnarray}
    
    \item the {\bf agent-preference} of $\mathcal{M}$ over $\mathcal{M}'$ by
    \begin{eqnarray}
        \mathcal{M} \underset{A}{\geq} \mathcal{M}' & \Longleftrightarrow\nonumber \\
        \forall a \in A,\ \mathcal{M}(a) \underset{a}{>} \mathcal{M}'(a)~ \vee ~\mathcal{M}(a) = \mathcal{M}'(a)
    \end{eqnarray}

    \item the {\bf strong task-preference} of $\mathcal{M}$ over $\mathcal{M}'$ by
    \begin{eqnarray}
        \mathcal{M} \underset{T}{>} \mathcal{M}' & \Longleftrightarrow \nonumber \\
         \mathcal{M} \underset{T}{\geq} \mathcal{M}'~ \wedge ~ \exists t \in T,\ \mathcal{M}(t) \underset{t}{>} \mathcal{M}'(t)
    \end{eqnarray}

    \item the {\bf strong agent-preference} of $\mathcal{M}$ over $\mathcal{M}'$ by
    \begin{eqnarray}
        \mathcal{M} \underset{A}{>} \mathcal{M}' & \Longleftrightarrow \nonumber \\ \mathcal{M} \underset{A}{\geq} \mathcal{M}'~ \wedge ~ \exists a \in A,\ \mathcal{M}(a) \underset{a}{>} \mathcal{M}'(a)
    \end{eqnarray}
    \end{enumerate}
%\end{definition}

In plain terms, matching $\mathcal{M}$ is task-preferred over $\mathcal{M}'$ if every task $t\in T$ prefers its match in $\mathcal{M}$ to its match in $\mathcal{M}'$ or it is assigned the same match. Strong task-preference further implies the existence of at least one task that prefers its match in $\mathcal{M}$ to its match in $\mathcal{M}'$, and therefore, matchings $\mathcal{M}$ and $\mathcal{M}'$ are different. A similar interpretation holds for the agent-based preferences.

Matching domination has played a very prominent role in the analysis of the {\em SM\/} and the {\em HR\/} stable-matching problems. In those cases, these relations define a distributive lattice structure over the set of stable matchings, that further guarantees the non-emptyness of this set and enables the design of efficient algorithms for its systematic exploration. However, as we shall see in the next section, the sable-matching problems considered in this work do not avail of any such structure.

\section{Matching Stability}
\label{sec:stab}

As pointed out in Section~\ref{sec:intro}, in the considered operational setting, a contemplated matching $\mathcal{M}$ is a tentative proposition that can be undermined by a ``blocking coalition'' of agents and tasks that can reassign themselves to each other in a more preferred manner for all of them. Hence, we are interested in ``stable'' matchings that are impervious to such undermining. Next, we formalize these concepts and requirements for the matching problem considered in this work.

%In stable-matching theory, it is assumed that any specified matching $\mathcal{M}$ is a tentative proposition, and it does not have a binding power over the matched parties. More specifically, matching $\mathcal{M}$ can be undermined {\em unilaterally\/} by any party that prefers to remain unassigned or maybe to be assigned only a part of its allocation specified by $\mathcal{M}$. In addition, a group of the involved parties may cooperate in an effort to reallocate themselves to each other in a way that results to more preferred allocations for each member of the group;\footnote{and, at the same time, might harm some parties that do not constitute part of the group} such a group is known as a {\em ``blocking coalition''\/} in the relevant terminology. A matching $\mathcal{M}$ is {\em stable\/} if it is impervious to unilateral and coalition-based undermining. 

%\subsection{Definitions and Properties}
%\label{ssec:d-p}

{\bf Matching stability:} We start the discussion of this section with a definition that formalizes the notion of the ``choice'' of an agent $a\in A$ over an offered set of tasks, $S$.
\begin{defi}
\label{defi:ch}
For any agent $a\in A$ and a set $S\subseteq T$, let
\begin{equation}
F(a,S)\ =\ \big \{S' \subseteq S:\ S'\in F(a) \big \}
\label{eq:feas-a}
\end{equation}
The {\bf choice of agent} $a$ {\bf over} $S$ is defined by
\begin{equation}
Ch(a,S)\ =\ \hat{S}\in F(a,S)\ \mbox{s.t.}\ \forall S'\in F(a,S)\setminus\{\hat{S}\},\  \hat{S} \underset{a}{>} S'
\label{eq:ch-a}
\end{equation}
Similarly, for any task $t\in T$ and a set $S\subseteq A$, let
\begin{equation}
A(t,S)\ =\ \big\{ \{a\}:\ a\in S\cap A(t) \big \} \cup \big \{ \emptyset \big \}
\label{eq:feas-t}
\end{equation}
The {\bf choice of task} $t$ {\bf over} $S$ is defined by
\begin{equation}
Ch(t,S)\ =\ \hat{S}\in A(t,S)\ \mbox{s.t.}\ \forall S'\in A(t,S)\setminus\{\hat{S}\},\  \hat{S} \underset{t}{>} S'
\label{eq:ch-a}
\end{equation}
\end{defi}

In general, it is conceivable that a matching ${\mathcal M}$ may be undermined {\em unilaterally\/} by a task or an agent, if they prefer to remain unassigned or, in the case of an agent, to be assigned only to a subset of its current match under ${\mathcal M}$.
The next definition and the ensuing proposition characterize the impossibility of such unilateral undermining of any given matching ${\mathcal M}$, by a single agent or task, in the considered problem setting.

\begin{defi}
\label{defi:u-st}
A matching ${\mathcal M}$ is {\bf individually rational} if 
\begin{equation}
\forall x\in A\cup T,\ \ Ch\big (x, {\mathcal M}(x)\big )\ =\ {\mathcal M}(x)
\label{eq:u-st}
\end{equation}
\end{defi}

\begin{prop}
\label{prop:u-st}
Every matching ${\mathcal M} \in \mathbb{M}$ is individually rational.
\end{prop}

The validity of Proposition~\ref{prop:u-st} results immediately from (i) Equations~\ref{eq:ab-1} and~\ref{eq:ab-2}, and (ii) the lexicographic ordering of the preference lists $P(a)$ for all agents $a\in A$.
%{\em Proof:} The validity of Proposition~\ref{prop:u-st} for tasks $t\in T$, results from the fact that ${\mathcal M}(t) \subseteq A(t)$ and Equation~\ref{eq:ab-1}. Similarly, the validity of Proposition~\ref{prop:u-st} for agents $a\in A$, results from the fact that ${\mathcal M}(a) \in F(a)$, the downward closure of $F(a)$, and the lexicographic ordering of the subsets of ${\mathcal M}(a)$ in agent $a$'s preference list $P(a)$. \hfill $\Box$
%Proposition~\ref{prop:u-st} guarantees that, in the considered problem setting, no task or agent will undermine unilaterally any matching $\mathcal{M}\in \mathbb{M}$.
Next, we characterize the blocking coalitions for a matching ${\mathcal M} \in \mathbb{M}$.\footnote{
Definition~\ref{defi:b-c} adapts to the considered matching problem a more general notion of subverting coalitions that are formed in the context of cooperative games. Matchings $\mathcal{M}\in \mathbb{M}$ are the possible outcomes of this game, and stable matchings constitute a {\em ``core''\/} of ``undominated'' outcomes. We refer the reader to \cite{RothSot90} for further details on the connection between the theory of cooperative games and stable-matching theory.}

\begin{defi} \label{defi:b-c}
    A \textbf{coalition} $\mathcal{C}$ is a subset of agents and tasks with $\mathcal{C} \cap A \ne \emptyset ~\wedge~ \mathcal{C} \cap T \ne \emptyset$.
Coalition $\mathcal{C}$ is \textbf{blocking} for a matching $\mathcal{M}$ if there exists a matching $\mathcal{M}'$ such that
    \begin{itemize}
        \item[(i)] $\forall t\in \mathcal{C},\ \mathcal{M}'(t) \in \mathcal{C}$;
       
        \item[(ii)]  $\forall t' \in T \backslash \mathcal{C},\  \mathcal{M}'(t') = \mathcal{M}(t')\ \vee\ \mathcal{M}'(t') = \emptyset$;
        
        \item[(iii)]  $\forall x\in \mathcal{C},\ \ \mathcal{M}'(x)\ \underset{x}{>}\ \mathcal{M}(x)$.
    \end{itemize}  
 \end{defi}

 Some further implications of Definition~\ref{defi:b-c}, that also provide an intuitive explanation for it, are as follows:
 \begin{enumerate}
     \item Parts (i) and (iii) of the definition imply that every task $t\in \mathcal{C}$ either is unassigned in matching $\mathcal{M}$ or it is assigned in $\mathcal{M}$ but it gets a better match in $\mathcal{M'}$.
     \item Parts (i) and (ii) of the definition imply that for every agent $a\in \mathcal{C},\ \mathcal{M'}(a)\setminus \mathcal{M}(a)\ \subset\ \mathcal{C}$; i.e., the new tasks allocated to an agent $a\in \mathcal{C}$ in matching $\mathcal{M'}$ with respect to its allocation in matching $\mathcal{M}$ belong in the coalition $\mathcal{C}$. Part (iii) of the definition also ensures that the allocation of these new tasks to agent $a$ results in a match $\mathcal{M'}(a)$ for this agent in $\mathcal{M'}$ that is preferred to its match $\mathcal{M}(a)$ in $\mathcal{M}$.
     \item On the other hand, in order to accommodate its newly allocated tasks in a feasible manner, an agent $a\in \mathcal{C}$ might have to drop some of the tasks in $\mathcal{M}(a)\setminus \mathcal{C}$. The dropped tasks will be unassigned in matching $\mathcal{M'}$. 
 \end{enumerate}

 \begin{defi} \label{def: stable}
    A matching $\mathcal{M}$ is \textbf{stable} if it is not blocked by any coalition. 
\end{defi}

The set of stable matchings is denoted by $\mathbb{M}_S$. We also denote the set of \textbf{unstable} matchings by $\mathbb{M}_U = \mathbb{M}\setminus \mathbb{M}_S$.

{\bf Blocking pairs:} A coalition $\mathcal{C}$ is a \textbf{pair} if $ |\mathcal{C}|=2$. Hence, a pair consists of a single task $t\in T$ and a single agent $a\in A$. In the following, we shall denote such a pair by ${\mathcal Q}$.

The next proposition establishes that we can assess the stability of any matching $\mathcal{M}\in \mathbb{M}$ by considering only the existence of a {\bf blocking pair\/}.

%\begin{defi} \label{def:stable-pairs}
%    For each agent $a\in A$ and a feasible allocation $S_i^a\in F(a)$, the pair $(a,S_i^a)$ is \textbf{stable} if there exists a matching $\mathcal{M}\in \mathbb{M}_S$ with $\mathcal{M}(a)=S^a_i$.
%\end{defi}

 \begin{prop}\label{prop: c-p}
     Under the lexicographic preferences $P(a)$ for the feasible allocations of every agent $a\in A$, every blocking coalition $\mathcal{C}$ for some matching $\mathcal{M}\in \mathbb{M}_U$ contains a blocking pair $\mathcal{Q}$.
 \end{prop}

{\em Proof:}
    Consider an unstable matching $\mathcal{M}\in \mathbb{M}_U$ with a blocking coalition $\mathcal{C}$ and the corresponding matching $\mathcal{M}'$ of Definition~\ref{defi:b-c}. Also, pick a task $\hat{t} \in {\cal C}$; this task exists since $\mathcal{C} \cap T \ne \emptyset$. According to Definition~\ref{defi:b-c}, there exists  an agent $\hat{a}$ in $\mathcal{C}$  such that 
    \begin{equation}
        \mathcal{M'}(\hat{t})\ =\ \hat{a}\ \underset{\hat{t}}{>}\ \mathcal{M}(\hat{t})
    \label{eq:bc-1}
    \end{equation}
    
 Equation~\ref{eq:bc-1} further implies that $\hat{t} \in \mathcal{M}'(\hat{a})$. More specifically, there is   
some $k \geq 0$  such that
     \begin{equation}
        \mathcal{M}(\hat{a}) = t_1\ldots t_k\ \vee\  \mathcal{M}(\hat{a}) = t_1\ldots t_k t_{k+1} \ldots
    \label{eq:bc-2}
    \end{equation}   
    \begin{equation}
        \mathcal{M}'(\hat{a}) = t_1\dots t_k \hat{t} \ldots
    \label{eq:bc-3}
    \end{equation}
    \begin{equation}
        \hat{t}\ \underset{\hat{a}}{>}\ t_{k+1}\ \underset{\hat{a}}{>}\ \emptyset
    \label{eq:bc-4}
    \end{equation}

Next we show that the pair $\mathcal{Q} = \{\hat{t},\hat{a}\}$ is a blocking pair for matching $\mathcal{M}$. For this, let us consider the assignment $\mathcal{M''}$ that is defined as follows:
\begin{equation}
    \mathcal{M''}\ = \ \big \{ (\hat{a}, t):\ \forall t\in \{t_1,\ldots,t_k, \hat{t}\} \big \}
    \label{eq:M''}
\end{equation}
The tasks $t_1,\ldots,t_k$ appearing in the above equation are those appearing in Equations~\ref{eq:bc-2}--\ref{eq:bc-3}. The definition of ${\mathcal M}''$ by Equation~\ref{eq:M''} further implies that
\begin{equation}
    \forall t\in T,\ \ |\mathcal{M''}(t)|\ \leq\ 1
    \label{eq:impl-1}
\end{equation}
\begin{equation}
    \forall a\in A,\ \ \mathcal{M''}(a)\ =\ \left\{ 
    \begin{tabular}{ll}
       $t_1\ldots t_k\hat{t}$  &  if $a=\hat{a}$ \\
        $\emptyset$ & o.w.
    \end{tabular} \right .
    \label{eq:impl-2}
\end{equation}
Also, since $\mathcal{M'}$ is a matching, $\mathcal{M}'(\hat{a}) \in F(\hat{a})$. But then, Equation~\ref{eq:bc-3} together with the downward closure of the sets $F(a)$, imply that 
\begin{equation}
    \mathcal{M''}(\hat{a})\in F(\hat{a})
    \label{eq:impl-3}
\end{equation}
Equations~\ref{eq:impl-1}--\ref{eq:impl-3} collectively imply that assignment $\mathcal{M}''$ is actually a matching, i.e., 
\begin{equation}
    \mathcal{M''}\in \mathbb{M}
    \label{eq:impl-4}
\end{equation}
The proof concludes by showing that matching $\mathcal{M''}$ satisfies the conditions of Definition~\ref{defi:b-c} with respect to the original matching $\mathcal{M}$ and the considered pair $\mathcal{Q}$.
Indeed, matching $\mathcal{M''}$ satisfies Condition (i) of Definition~\ref{defi:b-c} thanks to Equation~\ref{eq:M''}. Equation~\ref{eq:M''} also implies that matching $\mathcal{M''}$ satisfies Condition (ii) of Definition~\ref{defi:b-c}. Condition (iii) of Definition~\ref{defi:b-c} holds for matching $\mathcal{M''}$ and the task $\hat{t}$ thanks to Equation~\ref{eq:bc-1}. 
Finally, Condition (iii) of Definition~\ref{defi:b-c} holds for matching $\mathcal{M''}$ and the agent $\hat{a}$ thanks to (a) Equations~\ref{eq:bc-2}--\ref{eq:bc-4}, (b) the working assumptions regarding the roles of agent $\hat{a}$, and the matchings $\mathcal{M}$ and $\mathcal{M'}$, (c) Equation~\ref{eq:impl-3}, and (d) the lexicographic preferences of agent $\hat{a}$ with respect to the task list $T(\hat{a})$.
\hfill $\Box$

The next corollary follows immediately from Proposition~\ref{prop: c-p}

 \begin{cor}\label{cor:stable}
     A matching $\mathcal{M}\in \mathbb{M}$ is stable if there is no blocking pair $\mathcal{Q}=\{a,t\}$ for it.
 \end{cor}

{\bf Assessing matching stability:}
Corollary~\ref{cor:stable} suggests an efficient algorithm for checking the stability of any given matching $\mathcal{M}\in \mathbb{M}$. This algorithm will iteratively search for a pair $(a,t) \in {\mathcal E}\setminus {\mathcal M}$ such that 
\begin{equation}
a\in Ch\big (t,\mathcal{M}(t)\cup \{a\}\big )\ \wedge\ t\in Ch\big (a,\mathcal{M}(a)\cup \{t\}\big )
\label{eq:st-ch}
\end{equation}
and it will terminate with a positive outcome only if such a pair is not found. The assessment of the second part of the condition of Equation~\ref{eq:st-ch} can be performed efficiently by (i) ordering the set $\mathcal{M}(a)\cup \{t\}$ in decreasing preference of its elements by agent $a$ to obtain a task string $s$, and (ii) assessing the feasibility of the prefix $\hat{s}$ of $s$ that is defined by task $t$, with respect to the feasibility set $F(a)$. 
Due to the imposed space limitations, we leave the relevant implementational details to the reader, but it is clear that, under the assumption of Footnote~\ref{foot:oracle}, the resulting algorithm is of polynomial complexity with respect to $|A|$ and $|T|$.

 {\bf The structure of $\mathbb{M}_S$:} 
 Next, we provide two examples showing, respectively, that (i) $\mathbb{M}_S = \emptyset$ for some problem instances, and (ii) when $\mathbb{M}_S \not= \emptyset$, the number of the matched tasks by two stable matchings $\mathcal{M}$ and $\mathcal{M}'$ may be different.
 
\begin{ex}
\label{ex:empty}
Let us consider a problem instance where $T=\{t_1,t_2,t_3\}$ and $A=\{a_1,a_2\}$. The task preferences over the set of agents, $A$, are as follows:\footnote{Notice that the pair $(a_2,t_3) \not\in {\mathcal E}$.} 
    \begin{equation*}
        t_1:\ a_2a_1 ~~~~~~ t_2:\ a_1a_2 ~~~~~~ t_3:\ a_1
    \end{equation*}
The agent preferences are represented by the lexicographic trees $\mathcal{D}(a_1)$ and $\mathcal{D}(a_2)$ depicted in Figure~\ref{D-ex_noSM}. 

\begin{figure}[t]
    \centering
      \begin{tikzpicture}[level/.style={sibling distance=20mm/#1}]
\node [circle,draw] (z){$ $}
  child {node [circle,draw] (a) {$t_1$}
    child {node [circle,draw] (a1) {$t_2$}}}
  child {node [circle,draw] (b) {$t_3$}}
  child {node [circle,draw] (c) {$t_2$}
  child [grow=right] {node (1) {Level 1} edge from parent[draw=none]
          child [grow=up] {node[rectangle,draw]  (0) {\textbf{$\mathcal{D}(a_1)$}} edge from parent[draw=none]
            child [grow=down] {node (nn) {$ $} edge from parent[draw=none]
            child [grow=down] {node (2) {Level 2} edge from parent[draw=none]}}
          }
    }} 
  ;
\end{tikzpicture}
%\hspace{1cm}
\vskip 0.1in
\begin{tikzpicture}[level/.style={sibling distance=20mm/#1}]
\node [circle,draw] (z){$ $}
  child {node [circle,draw] (a) {$t_2$}}
  child {node [circle,draw] (b) {$t_1$}
  child [grow=right] {node (1) {Level 1} edge from parent[draw=none]
          child [grow=up] {node[rectangle,draw]  (0) {\textbf{$\mathcal{D}(a_2)$}} edge from parent[draw=none]
            child [grow=down] {node (nn) {$ $} edge from parent[draw=none]
            child [grow=down] {node (2) {$ $} edge from parent[draw=none]}}
          }
    }} 
  ;
\end{tikzpicture}

\vskip -0.45in
\caption{The lexicographic trees $\mathcal{D}(a_1)$ and $\mathcal{D}(a_2)$ in Example~\ref{ex:empty}.}
\label{D-ex_noSM}
\end{figure}
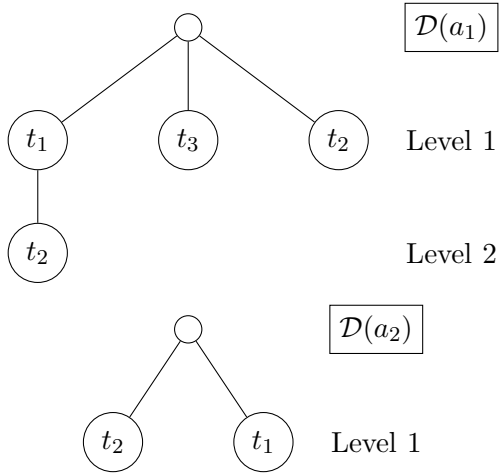

\begin{table}[t]
\centering
\begin{tabular}{|c||c|c|c|c}
\cline{1-4}
   $a_1\backslash a_2$         & $t_2$ & $t_1$ & $\emptyset$ &  \\  \cline{1-4} \noalign{\vskip\doublerulesep
         \vskip-\arrayrulewidth}  \cline{1-4}
$t_1t_2$    &  $\times$     &   $\times$    &    $(t_1,a_2)$         &  \\ \cline{1-4}
$t_1$       &    $(t_2,a_1)$   &  $\times$     &   $(t_2,a_2)$          &  \\ \cline{1-4}
$t_3$       &  $(t_1,a_1)$     &  $(t_2,a_2)$     &   $(t_1,a_2)$          &  \\ \cline{1-4}
$t_2$       &  $\times$     &  $(t_3,a_1)$     &     $(t_1,a_2)$        &  \\ \cline{1-4}
$\emptyset$ &   $(t_1,a_1)$    &   $(t_2,a_2)$    &     $(t_1,a_1)$        &  \\ \cline{1-4}
\end{tabular}

\caption{Matching-stability analysis for Example~\ref{ex:empty} }
\label{table:block}
\end{table}

For this example, we assessed the stability of every possible matching $\mathcal{M} \in \mathbb{M}$, and the results are reported in Table~\ref{table:block}. 
The first column of Table~\ref{table:block} enumerates all the feasible allocations to agent $a_1$, as specified by tree $\mathcal{D}(a_1)$. Similarly, the first row of Table~\ref{table:block} enumerates all the feasible allocations to agent $a_2$, according to tree $\mathcal{D}(a_2)$. Every other cell of Table~\ref{table:block} represents an assignment ${\cal A}$ that is defined by the allocations to agents $a_1$ and $a_2$ that correspond to this cell.

Cells marked by `$\times$' correspond to assignments that do not constitute valid matchings, since some task is assigned to more than one agent. 
Every other cell corresponds to a valid matching, but the cell also reports a blocking pair for this matching. Therefore, for the matching problem instance considered in this example, $\mathbb{M}_S = \emptyset$. \footnote{
This example is an adaptation to the considered problem setting of a similar example presented in \cite{mcdermid2010keeping} for establishing the potential emptiness of the stable-matching set for HRS problem instances. Also, since, according to Remark~\ref{rem:HRS}, the considered stable-matching problem subsumes the {\em HRS\/} stable-matching problem, the decision problem of resolving the non-emptiness of $\mathbb{M}_S$, for any given instance of this problem, is NP-complete.}
\end{ex}

\begin{figure}[t]
    \centering
      \begin{tikzpicture}[level/.style={sibling distance=20mm/#1}]
\node [circle,draw] (z){$ $}
  child {node [circle,draw] (a) {$t_1$}}
  child {node [circle,draw] (b) {$t_2$}
        child {node [circle,draw] (b1) {$t_3$}}}
  child {node [circle,draw] (c) {$t_3$}
  child [grow=right] {node (1) {Level 1} edge from parent[draw=none]
          child [grow=up] {node[rectangle,draw]  (0) {\textbf{$\mathcal{D}(a_1)$}} edge from parent[draw=none]
            child [grow=down] {node (nn) {$ $} edge from parent[draw=none]
            child [grow=down] {node (2) {Level 2} edge from parent[draw=none]}}
          }
    }} 
  ;
\end{tikzpicture}
%\hspace{1cm}
\vskip 0.1in
\begin{tikzpicture}[level/.style={sibling distance=20mm/#1}]
\node [circle,draw] (z){$ $}
  child {node [circle,draw] (a) {$t_2$}}
  child {node [circle,draw] (b) {$t_1$}
  child [grow=right] {node (1) {Level 1} edge from parent[draw=none]
          child [grow=up] {node[rectangle,draw]  (0) {\textbf{$\mathcal{D}(a_2)$}} edge from parent[draw=none]
            child [grow=down] {node (nn) {$ $} edge from parent[draw=none]
            child [grow=down] {node (2) {$ $} edge from parent[draw=none]}}
          }
    }} 
  ;
\end{tikzpicture}
\vskip -0.45in
\caption{The lexicographic trees $\mathcal{D}(a_1)$ and $\mathcal{D}(a_2)$ in Example~\ref{ex:uneq}.}
\label{D-ex_noSM3}
\end{figure}
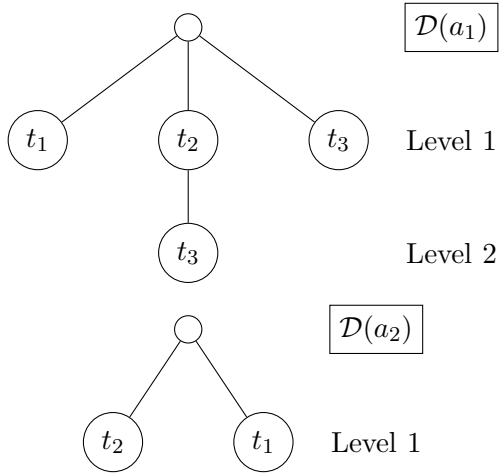

 \begin{ex}
 \label{ex:uneq}
 In this example we consider a problem instance with task set $T=\{t_1,t_2,t_3\}$ and agent set $A=\{a_1,a_2\}$. The task preferences over $A$ are:
    \begin{equation*}
        t_1:\ a_2a_1 ~~~~~~ t_2:\ a_1a_2 ~~~~~~ t_3:\ a_1
    \end{equation*}

The agent preferences are represented by the lexicographic trees $\mathcal{D}(a_1)$ and $\mathcal{D}(a_2)$ depicted in Figure~\ref{D-ex_noSM3}.

It can be checked that the matchings $\mathcal{M}_1$ and $\mathcal{M}_2$ defined by
\begin{equation}
\mathcal{M}_1(a_1)=t_1~~~~~~~\mathcal{M}_1(a_2)=t_2
\label{eq:comp-1}
\end{equation}
and
\begin{equation}
\mathcal{M}_2(a_1)=t_2 t_3~~~~~~~\mathcal{M}_2(a_2)=t_1
\label{eq:comp-2}
\end{equation}
are stable. Furthermore, task $t_3$ is assigned in $\mathcal{M}_2$ but it is not assigned in $\mathcal{M}_1$. Hence, the ``Rural Hospitals'' theorem does not hold for the considered class of stable-matching problems, and the maximization of the number of assigned tasks in a stable manner, or of a more general function of the corresponding task sets, becomes a meaningful optimization problem.
 \end{ex}

\section{An IP-based representation of matching stability and optimal stable matchings}
\label{sec:IP}

In this section we provide an algebraic representation of the space of stable matchings, $\mathbb{M}_S$, for any given problem instance, and we show how we can use this representation for a systematic exploration of $\mathbb{M}_S$ for preferred stable matchings by means of IP formulations. We also address briefly some computational challenges of the presented method.

The presented developments are similar, in spirit, to the past LP and IP-based formulations of the {\em SM} problem and the various versions of the ${HR\/}$ problems \cite{Manlovebook}. But  in the considered problem context, the derivation of the sought representation of $\mathbb{M}_S$ is further complicated by the arbitrary structure of the agent preference lists $P(a)$, and it relies heavily on the representation of these lists by the corresponding lexicographic trees ${\mathcal D}(a)$.

{\bf The proposed representation of $\mathbb{M}_S$:}
Our algebraic representation of the set of stable matchings, $\mathbb{M}_S$, 
takes the form of a set of {\em linear\/} constraints on a set of {\em binary\/} variables that represent the considered allocations. These binary variables are defined as follows:

\begin{equation}
    \forall a \in A,\ \forall n \in \mathcal{D}(a),\ \ \ X_{an}  = \mathbb{I}_{\{a\text{ is assigned task set } T(n)\}}
    \label{eq:d-v}
\end{equation}

In Equation~\ref{eq:d-v}, $n$ denotes a node of the lexicographic tree $\mathcal{D}(a)$ and $T(n)$ is the task allocation for agent $a$ corresponding to node $n$. The set of stable matchings, $\mathbb{M}_S$, is represented by the valuations of the variable set $\{X_{an}: a\in A,\ n\in {\mathcal D}(a)\}$ that satisfy the following constraints:
\begin{equation}
 \forall a \in A,\ \ \underset{n \in \mathcal{D}(a)}{\sum} X_{an} = 1
 \label{eq:c-1}
 \end{equation}
 \begin{equation}
\forall t \in T,\ \ \underset{a \in A}{\sum}~\underset{\underset{t \in T(n)}{n \in \mathcal{D}(a):}}{\sum} X_{an}                                                            \leq 1 
\label{eq:c-2}
\end{equation}
\begin{eqnarray}
\forall t \in T,\ \forall a \in A(t),\ \ \underset{\underset{\underset{\underset{\text{has a child } n'' \text{ with } t \in T(n'')}{T(n')=\{t' \in T(n) : t' \underset{a}{>} t\}}}{t \notin T(n) \wedge \text{ node $n'$ with}}}{n \in \mathcal{D}(a):}}{\sum} X_{an}  & \leq \nonumber \\
\underset{\underset{a' \underset{t}{\geq} a}{a' \in A(t)}}{\sum}~\underset{\underset{t \in T(n)}{n \in \mathcal{D}(a):}}{\sum} X_{a'n}  
\label{eq:c-3}
\end{eqnarray}
\begin{equation}
\forall a \in A,\ \forall n \in \mathcal{D}(a),\ \ X_{an}\  \in\ \{0,1\}
\label{eq:c-4}
\end{equation}

Constraint~\ref{eq:c-1} expresses the requirement that each agent $a\in A$ must be assigned one and only one task allocation among the allocations that are represented by the nodes $n \in {\mathcal D}(a)$.\footnote{We remind the reader that the root node $o$ of ${\mathcal D}(a)$ represents the possibility that agent $a$ is not assigned any tasks.} Constraint~\ref{eq:c-2} expresses the requirement that a task $t \in T$ is assigned to at most one agent.
Constraint~\ref{eq:c-4} enforces the binary nature of the employed decision variables $X_{an}$. Collectively, these three constraints ensure that any satisfying pricing of the variables $X_{an}$ specifies a valid matching ${\mathcal M} \in \mathbb{M}$. For further reference, let us denote any such matching by ${\mathcal M}[X_{an}]$.

Constraint~\ref{eq:c-3} expresses the stability requirement for the matchings ${\mathcal M}[X_{an}]$.
More specifically, the instantiation of this constraint for any particular pair $(t,a) \in T\times A(t)$, ensures that the pair $(t,a)$ is not a blocking pair for ${\mathcal M}[X_{an}]$.

In order to explain the logic encoded by Constraint~\ref{eq:c-3}, first we notice that every pricing of the variables $X_{an}$ that satisfies the other three constraints, will result in a value of either 0 or 1 for each of the two sides of Constraint~\ref{eq:c-3}. More specifically, a value of 1 for the right-hand-side of the instantiation of Constraint~\ref{eq:c-3} for some pair $(t,a)\in T\times A(t)$,  implies that, in the corresponding matching ${\mathcal M}[X_{an}]$, task $t$ is assigned either to agent $a$ or to some agent $a'$ that it prefers to $a$. On the other hand, a value of 1 for the left-hand-side of the considered instantiation of Constraint~\ref{eq:c-3} under matching 
${\mathcal M}[X_{an}]$, implies that $t\not \in {\mathcal M}[X_{an}](a)\ \wedge\ t\in Ch\big (a, {\mathcal M}[X_{an}](a)\cup \{t\}\big )$. Hence, if the right-hand-side of the considered constraint was equal to 0 and the left-hand-side was equal to 1, for some matching ${\mathcal M}[X_{an}]$, task $t$ and agent $a$ would constitute a blocking pair for matching ${\mathcal M}[X_{an}]$, and this matching would be unstable. For any other pricing of the left- and the right-hand-sides of the considered constraint, the pair $(t,a)$ is non-blocking for the corresponding matching ${\mathcal M}[X_{an}]$, since either task $t$ is assigned favorably with respect to $a$ and/or agent $a$ would not choose task $t$ if it was offered to it.
Hence, Constraint~\ref{eq:c-3} prevents exactly those pricings of $X_{an}$ that might lead to a valid but unstable matching $\mathcal{M}[X_{an}]$. 

{\bf Searching for ``optimal'' stable matchings through IP:}
The availability of the Constraint set~\ref{eq:c-1}--\ref{eq:c-4} enables the implicit search for {\em ``optimal''\/} stable matchings ${\mathcal M}\in \mathbb{M}_S$ with respect to any objective function that is additive to the values defined by the agent allocations to the task sets that are represented by the variables $X_{an}$, through the formulation and solution of a linear IP. More specifically, we consider the case where the allocation $X_{an}, a\in A,\ n\in {\mathcal D}(a)$, incurs a value of $c_{an}$, and we want to maximize the total value $\sum_{a\in A} \sum_{n\in \mathcal{D}(a)} c_{an} X_{an}$.
The corresponding formulation is a binary program and it can be solved, in principle, through commercial solvers.
The next example motivates more concretely the need for these formulations and 
highlights the structure of Constraints~\ref{eq:c-1}--\ref{eq:c-4}.

\begin{ex}
\label{ex:form}
As established in Example~\ref{ex:uneq}, different stable matchings in the set $\mathbb{M}_S$ may assign different subsets of the task set $T$ to the available agents, and these subsets might even differ in their cardinality. Hence, one can consider finding a stable matching ${\mathcal M}\in \mathbb{M}$ that maximizes the total number of the allocated tasks across all agents $a\in A$. In this case, $\forall a\in A,\ \forall n\in {\mathcal D}(a),\ c_{an} = |T(n)| \geq 0$. \footnote{This objective may be important, for instance, to clearing houses that receive a fixed premium for every matched task; c.f.\ the application example outlined in Section~\ref{sec:intro}. But the objective also holds considerable theoretical interest since it relates to the notion of {\em maximal\/} matchings with respect to the ``task-preference'' ordering of $\mathbb{M}_S$ that was defined in Section~\ref{sec:m-prob}.}

For the problem instance of Example~\ref{ex:l-t}, the above optimization problem leads to the following IP formulation:

$$\max\ \ \ 2X_{11} + X_{12} + X_{13} + X_{21} + X_{22}$$
s.t.
$$ X_{11} + X_{12} + X_{13} + X_{14}  = 1$$
 $$ X_{21} + X_{22} + X_{23}  = 1 $$
 $$ X_{11} + X_{12} + X_{22} \leq 1$$
 $$ X_{11} + X_{13} + X_{21} \leq 1 $$
 $$ X_{13} + X_{14}  \leq X_{11} + X_{12} + X_{22}$$
 $$ X_{23}  \leq  X_{22} $$
 $$ X_{12} + X_{14}  \leq X_{11} + X_{13} $$
 $$ X_{22} + X_{23} \leq X_{11} + X_{13} + X_{21} $$
 $$ X_{11}, X_{12}, X_{13}, X_{14}, X_{21}, X_{22}, X_{23}  \in \{0,1\} $$

It can be easily verified that the only feasible solution for this IP is $X_{13}=X_{22}=1$ and every other variable $X_{an} = 0$. The corresponding objective value is equal to 2.
\end{ex}

{\bf Coping with the potential emptiness of $\mathbb{M}_S$:}
When $\mathbb{M}_S = \emptyset$, the IPs formulated in the previous paragraph will be infeasible. To cope with this issue, we may allow the presence of some blocking pairs in the constructed matching ${\mathcal M}$. But we stipulate that the tolerated instability must be the minimum necessary for obtaining a feasible solution.
This logic can be brought in the above IP formulation by introducing the following additional set of binary variables:
\begin{equation}
\forall t \in T,\ \forall a \in A(t),\ \ Y_{ta} = \mathbb{I}_{\{\text{pair } (t,a) \text{ is blocking}\}}
\label{eq:r-v}
\end{equation}

Variables $Y_{ta}$ are used to {\em relax\/} the stability-enforcing Constraint~\ref{eq:c-3} through its substitution with the following constraint:
\begin{eqnarray}
\forall t \in T,\ \forall a \in A(t),\ \ \underset{\underset{\underset{\underset{\text{has a child } n'' \text{ with } t \in T(n'')}{T(n')=\{t' \in T(n) : t' \underset{a}{>} t\}}}{t \notin T(n) \wedge \text{ node $n'$ with}}}{n \in \mathcal{D}(a):}}{\sum} X_{an}  & \leq \nonumber \\
\underset{\underset{a' \underset{t}{\geq} a}{a' \in A(t)}}{\sum}~\underset{\underset{t \in T(n)}{n \in \mathcal{D}(a):}}{\sum} X_{a'n} \ +\ Y_{ta}
\label{eq:c-3'}
\end{eqnarray}

From the provided explanation of the functionality of the original Constraint~\ref{eq:c-3}, it should be clear that the addition of the binary variable $Y_{ta}$ to the right-hand-side of this constraint renders the constraint satisfiable for any pricing of the variables $X_{an}$ by setting this variable equal to 1. 

The requirement that (i) a variable $Y_{ta}$ should be priced to `1' only when it is necessary in order to establish the IP feasibility, and (ii) in such a case, the number of these variables must be minimized, is enforced in the modified IP formulation by replacing the original objective function with the following one:
\begin{equation}
\sum_{a\in A} \sum_{n\in \mathcal{D}(a)} c_{an} X_{an} - c_y \cdot \sum_{t\in T} \sum_{a\in A(t)} Y_{ta}
\label{eq:n-obj}
\end{equation}

Furthermore, in the above expression, the ``cost'' coefficient $c_y$ must be set to an upper bound of $\sum_{a\in A} \sum_{n\in \mathcal{D}(a)} c_{an} X_{an}$ for all possible pricings of the variables $X_{an}$; in this way, the reduction of the sum $\sum_{t\in T} \sum_{a\in A(t)} Y_{ta}$ by one unit is more valuable than any possible improvement of the first part of the objective function.

{\bf Complexity considerations:}
It is easy to see from Equations~\ref{eq:c-1}--\ref{eq:c-3} that the number of the technological constraints in the considered IP  formulations is $O(|A||T|)$, i.e., polynomially related to the size of the primary entities that define the underlying matching problem. Furthermore, the binary nature of the decision variables employed by these formulations, together with the fact that all coefficients appearing in the constraints are equal to $\pm1$, would place these formulations among the easiest IP classes \cite{Wolsey}. A more intuitive and  practical explanation of this fact is obtained upon noticing that due the {\em logical\/} nature of (i) the semantics that define the employed variables and (ii) the dependencies that are expressed by the various constraints, 
 the pricing of a variable to one of its two possible values will also force many other variables to one of their two values. These dependencies can be efficiently traced and exploited by modern IP solvers, frequently in ``pre-processing'' stages that precede the algorithmic solution of the input formulations \cite{Acht+16}.

But the solution of the considered IPs is further complicated by the fact that the number of the employed decision variables $X_{an}$ is equal to $\sum_{a\in A} |{\mathcal D}(a)|$, and as pointed out in Section~\ref{sec:m-prob}, the sizes, $|{\mathcal D}(a)|$, of the lexicographic trees ${\mathcal D}(a)$ are super-polynomial functions of $|T|$, in general. This problem can be effectively addressed by applying a {\em column-generation\/} method \cite{Magn} that will bring the variables $X_{an}$ into the processed formulations in a controlled manner, based on their ability to improve the current incumbent solution.

A preliminary investigation with the  development and implementation of such a method for the considered IPs has indicated that the number of variables $X_{an}$ that will actually enter the explicitly solved formulations, is a small subset of their overall set. Furthermore, the search for new competitive variables $X_{an}$ to enter the ``master'' LP formulation at each iteration of the column-generation method can be parallelized over the subsets of these variables corresponding to each agent $a\in A$, while the structural information for each such variable subset and the underlying allocations that is provided by the corresponding lexicographic tree $\mathcal{D}(a)$ enables a very efficient organization of the corresponding search. More specifically, the semantics that define the tree $\mathcal{D}(a)$ enable (i) an incremental computation of the ``reduced costs'' of the variables $X_{an}$ -- i.e., the indices that define the competitiveness of these variables to enter the ``master''  LP -- across its different layers and paths, and (ii) the development of additional logic that can guide the search for such competitive variables within the tree.

Some additional effects that can control naturally the actual size -- especially the depth -- of the lexicographic trees  
$\mathcal{D}(a)$ and the empirical complexity of the search processes outlined above, may arise from (i) the notion of feasibility of the task allocations for the different agents employed in particular applications of the considered model, that might restrict practically the cardinality of these task sets to smaller values, and (ii) a potential tendency of the sought allocations to distribute rather evenly the available tasks among the agents.

Finally, it is also interesting to examine the possibility of the initialization of the aforementioned algorithms for the considered IPs by starting the ``master'' LP of the column-generation method with a pertinently selected set of variables. This set of variables might be obtained, for instance, by some heuristically constructed solution for the underlying matching problem, or even a solution to a relaxation of this problem obtained in the spirit of \cite{Dean+06,CsehDean16}. 
The imposed page limitation for this document does not allow a more expansive treatment of all the ideas and the techniques that were outlined in the previous paragraphs, and the corresponding developments are deferred to a follow-up publication.

\section{Substitutability}
\label{sec:sub}

The property of {\em ``substitutability''\/} characterizes ``special'' structure for preferences concerning selection between multi-item allocations by a single entity. Furthermore,
in the context of one-to-many and many-to-many stable-matching theory, this property has been associated with important structural properties of the stable-matching space, like (i) the non-emptiness of this space and (i) the existence of unique maximal elements with respect to each side's preferences that can be efficiently computed by {\em GS\/}-type of algorithms \cite{RothSot90}.

In the context of the considered stable-matching problem, substitutability pertains to the agent preferences over their feasible task allocations, and it is formally defined as follows:
\begin{defi}
 \label{defi:subst}
    Agent $a$'s preferences over its feasible task allocations in $F(a)$ have the property of \textbf{substitutability} if they satisfy the following condition:  For any task set $S\subseteq T$ with $|S| \geq 2$ and any two tasks $t$ and $t'$ contained in it, 
    
    \begin{equation}
    t\ \in\ Ch(a,S)\ \ \Longrightarrow\ \ t\ \in\ Ch(a,S\setminus \{t'\})
    \label{eq:subst-or}
    \end{equation}
\end{defi}

%where an agent expresses preferences over individual tasks as well as over sets of tasks, substitutability implies that when a task $t$ is in the preferred set of an agent $a$ that is called to select from a task set $S\subseteq T$, it remains in the preferred set of $a$ even when set $S$ is thinned of some of its other elements.
 Hence, substitutability {\em ``decouples''\/} the preference that is expressed by agent $a$ for any individual task $t$ from the overall composition of the entire subset of $S$ that is selected by this agent.\footnote{In economic-theoretic terms, substitutable preferences do not allow the presence of any notions of {\bf complementarity} among the elements of set $T$.} 

Next, we provide an alternative characterization of substitutability for our problem setting that relies on the representation of the preferences of any agent $a\in A$ for the task allocations in $F(a)$ by means of the lexicographic tree ${\cal D}(a)$. To formally state this result, we need the following notation: 
For any task set $S\subseteq T(a)$, let $\mathcal{D}(a;S)$ denote the subtree of $\mathcal{D}(a)$ consisting of all the maximal paths of $\mathcal{D}(a)$ that emanate from the root node $o$ and their nodes hold only tasks $t\in S$. Furthermore, for any node $n\in {\mathcal D}(a;S)$ with $n\not = o$, let ${\mathcal D}(n;a;S)$ denote the subtree of ${\cal D}(a;S)$ rooted at $n$. Also, let $l(n;S)$ denote the leftmost leaf node of the subtree $\mathcal{D}(n;a;S)$, and $\hat{n}$ denote the closest right sibling of $n$ in ${\mathcal D}(a;S)$. Then, we have the following proposition:

\begin{prop}
\label{prop:subs}
The preferences of an agent $a \in A$ for its feasible task allocations in $F(a)$ have the property of {\bf substitutability} if and only if, for all task sets $S\subseteq T(a)$ with $|S|\geq 2$, the corresponding subtree $\mathcal{D}(a;S)$ of
the lexicographic tree ${\mathcal D}(a)$ satisfies the following condition:
\begin{equation}
    \forall n \in \mathcal{D}(a;S),\ \ T\big (l(n;S)\big ) \setminus T(n)\ \subseteq\  T\big ( l(\hat{n};S) \big )
    \label{eq:subst}
\end{equation}
\end{prop}

{\em Proof:}
First we establish the necessity of the condition of Equation~\ref{eq:subst} for substitutability.
 Hence, consider a set $S\subseteq T(a)$ with $|S|\geq 2$ and a node $n$ in ${\mathcal D}(a;S)$ that violates the condition of Equation~\ref{eq:subst}.
 
Also, consider the subset $\hat{S}$ of $S$ defined by $\hat{S} = T\big ( l(n;S) \big )\cup T\big ( l(\hat{n};S) \big )$. By the lexicographic preferences of $a$, $Ch(a,\hat{S}) = T\big ( l(n;S) \big )$.

 Next, let $t'$ be the task held by node $n$, and notice that $Ch(a,\hat{S}\setminus \{t'\}) = T\big ( l(\hat{n};S) \big )$. Finally, pick a task $t$ in $\big (T \big ( l(n;S)\big ) \setminus  T(n) \big ) \setminus
 T\big ( l(\hat{n};S) \big )$; such a task exists by the working hypothesis for $n$. From the above definition of the task set $\hat{S}$ and the selection of the tasks $t$ and $t'$, $t$ and $t'$ belong in $\hat{S}$, $t$ also belongs in $Ch(a,\hat{S})$, but it does not belong in $Ch(a,\hat{S}\setminus \{t'\})$. 
Hence, the constructed triplet $\langle \hat{S}, t, t' \rangle$ violates the condition of Equation~\ref{eq:subst-or}.

Next, we establish the sufficiency of the condition of Equation~\ref{eq:subst} for substitutability, by contradiction. Hence, suppose that the condition of Proposition~\ref{prop:subs} holds, but there is a triplet $\langle S, t, t'\rangle$ violating the condition of Equation~\ref{eq:subst-or}.

Consider the nodes $n_1$ and $n_2$ of the lexicographic tree $\mathcal{D}(a)$ with $T(n_1) = Ch(a,S)$ and $T(n_2) = Ch(a,S\setminus \{t'\})$. Also, set $\hat{S} = Ch(a,S) \cup Ch(a,S\setminus \{t'\})$, and notice that $\mathcal{D}(a;\hat{S})$ consists of the two paths $\pi(n_1)$ and $\pi(n_2)$ of $\mathcal{D}(a)$. 

The lexicographic nature of agent $a$'s preferences together with the working hypothesis imply that tasks $t$ and $t'$ are held by some nodes of $\pi(n_1)$ but they do not appear on the path $\pi(n_2)$. Let $n$ denote the node of $\mathcal{D}(a)$ holding task $t'$ in $\pi(n_1)$, and $\tilde{n}$ denote the parent node of node $n$ in $\mathcal{D}(a)$. The lexicographic preferences of agent $a$ imply that the paths $\pi(n_1)$ and $\pi(n_2)$ have as a maximal common subpath the path $\pi(\tilde{n})$ leading from the root node $o$ to node $\tilde{n}$. Finally, let $\hat{n}$ denote the child of node $\tilde{n}$ on the path $\pi(n_2)$. 

The above construction implies that: (i) $\hat{n}$ is the unique right sibling of node $n$ in $\mathcal{D}(a;\hat{S})$; (ii) $T\big ((l(n;\hat{S}) \big ) = T(n_1) = Ch(a; S)$; and $T\big ((l(\hat{n};\hat{S}) \big ) = T(n_2) = Ch(a; S\setminus \{t'\})$. Furthermore, by the working hypothesis, the subpath $\pi(n,n_1)$ of path $\pi(n_1)$ leading from node $n$ to node $n_1$, holds the considered task $t$ that does not appear in the path $\pi(n_2)$. 

In the subtree $\mathcal{D}(a;\hat{S})$, the tasks held by the path $\pi(n,n_1)$ constitute the set $T \big (l(n;\hat{S})\big ) \setminus T(n)$, and the tasks held by the path $\pi(n_2)$ constitute the set $T\big ( l(\hat{n};\hat{S}) \big )$. Hence, according to the previous paragraph, task $t$ belongs in $T \big (l(n;\hat{S})\big ) \setminus T(n)$ but it does not belong in $T\big ((l(\hat{n};\hat{S}) \big )$; i.e.,  the constructed triplet $\langle \hat{S}, n, \hat{n}\rangle$ violates the condition of Equation~\ref{eq:subst}, which contradicts the working hypothesis. \hfill $\Box$

Proposition~\ref{prop:subs} provides an alternative tool for the analysis of the property of substitutability and its implications for the considered stable-matching problem. 
As an example of this possibility, and of the representational and analytical power that it brings in the corresponding investigations, we notice that Section 5 in \cite{mcdermid2010keeping} establishes that {\em HRS\/} problems where 
each hospital has at most two acceptable residents, are always feasible.
This result is established in \cite{mcdermid2010keeping} by providing a constructive algorithm for a stable matching. Proposition~\ref{prop:subs} enables a much shorter derivation of the aforementioned result  of \cite{mcdermid2010keeping}, and its immediate generalization to the stable-matching problem that is considered in this work, upon noticing that the lexicographic tree $\mathcal{D}(a)$ of any agent $a\in A$ with $|T(a)|\leq 2$ satisfies trivially the condition of Equation~\ref{eq:subst}; hence, the non-emptyness of the set $\mathbb{M}_S$ is obtained from the corresponding results of \cite{RothSot90} on substitutable preferences. It is also possible to redevelop, for the considered problem setting, the results of \cite{RothSot90} regarding the non-emptiness of $\mathbb{M}_S$ under substitutable preferences for the problem agents,  using the characterization of substitutability that is provided by Proposition~\ref{prop:subs} and the analytical characterization of the set $\mathbb{M}_S$ through the Constraints~\ref{eq:c-1}--\ref{eq:c-4}. Such a redevelopment would resemble, in spirit, the reproduction of the results on the {\em SM\/} problem that is provided in \cite{BalRat97}, based on the graph-theoretic representation of the {\em SM\/} problem that was introduced in that paper.

\section{Conclusions}
\label{sec:con}

Motivated by the increasing interest for an explicit representation and handling of ``preference'' structures in various internet-enabled business platforms, this work extended the past model of the {\em HR} stable-matching problem under the presence of resident ``sizes'', to a much broader modeling framework regarding the notion of the feasible allocations for the ``hospitals''. It provided a thorough formal characterization of this new stable-matching problem, important structural results for the space of stable matchings, an algebraic characterization of this space in the form of an integer program, and it also connected the pursued analysis to the notion of substitutability that is known to play a significant role in the well-posedness and the computational tractability of stable-matching problems with multi-item allocations.

Our future work on this problem will seek to fully develop the column-generation method that was outlined for the IP formulations of Section~\ref{sec:IP}, analyze the dynamics of the corresponding search process, and assess the  scalability of the method through numerical experimentation. Another interesting research direction is the adaptation of the relaxing approach of \cite{Dean+06,CsehDean16} for coping with infeasible cases in the context of the {\em HRS\/} problem, to the more general problem version that was introduced in this work, where the notions of ``congestion'' and ``over-allocation'' do not accept a closed-form characterization. Finally, we shall also seek to apply the more general theoretical developments to the task allocation that takes place in specific multi-agent-based service systems.

\bibliographystyle{IEEEtran}
\bibliography{SM,dlock}

\end{document}